\documentclass[preprint,prd,nofootinbib]{revtex4}

\usepackage{color} 
     
\usepackage{amssymb}
\usepackage{amsmath}
\usepackage{graphicx}

\usepackage{ulem}

\begin{document}

\title{Scalar field vacuum expectation value induced by gravitational wave background}
\author{Preston Jones}
\email{preston.jones1@erau.edu}

\affiliation{Embry Riddle Aeronautical University, Prescott, AZ 86301}

\author{Patrick McDougall}
\email{pmcdougall@mail.fresnostate.edu}

\affiliation{Department of Physics, California State University Fresno, Fresno, CA 93740}

\author{Michael Ragsdale}
\email{michael.ragsdale@fresnocitycollege.edu}

\affiliation{MSE Division, Fresno City College, Fresno, CA 93741}

\author{Douglas Singleton}
\email{dougs@csufresno.edu}

\affiliation{Department of Physics, California State University Fresno, Fresno, CA 93740 \\
and \\
Institute of Experimental and Theoretical Physics Al-Farabi KazNU, Almaty, 050040, Kazakhstan}

\date{\today}

\begin{abstract}
We show that a {\it massless} scalar field in a gravitational wave background can develop a non-zero vacuum expectation value. We draw comparisons to the generation of a non-zero vacuum expectation value for a scalar field in the Higgs mechanism and with the dynamical Casimir vacuum. We propose that this vacuum expectation value, generated by a gravitational wave, can be connected with particle production from gravitational waves and may have consequences for the early Universe where scalar fields are thought to play an important role.   
\end{abstract}

\maketitle

\section{Introduction}

The Brout-Englert-Higgs mechanism \cite{Higgs} is one of the cornerstones of the Standard Model of particle physics. Part of the Higgs mechanism involves a scalar field developing a non-zero vacuum expectation value rather than having a vacuum expectation value of zero. An example of this non-zero vacuum expectation value comes from $\Lambda \Phi ^4$ theory with a complex scalar field
whose Lagrangian density is
\begin{equation}
\label{higgs}
{\cal L} = \partial_\mu \Phi ^* \partial ^\mu \Phi - \frac{1}{2} m^2 | \Phi | ^2 - \frac{1}{4} \Lambda | \Phi |^4 ~.
\end{equation} 
The equation of motion from \eqref{higgs} is 
\begin{equation}
\label{higgs-em}
\partial _\mu \partial ^\mu \Phi - m^2 \Phi - \Lambda \Phi^3 = 0  ~.
\end{equation} 
If we look for solutions, $\Phi$, which are space and time independent ({\it i.e.} $\partial _\mu \Phi =0$) {\it and} if $m^2 >0$ then the only solution is $\Phi =0$. However for a tachyonic mass term ({\it i.e.} $m^2 <0$) \eqref{higgs-em} has a non-zero, constant solution $\Phi_0 = \langle 0 |\sqrt{\Phi ^* \Phi} | 0 \rangle = \sqrt{\frac{-m^2}{\Lambda}}$. The vacuum solution is now given by $\Phi = \sqrt{\frac{-m^2}{\Lambda}} e^{i \theta}$ with magnitude  $\sqrt{\frac{-m^2}{\Lambda}}$ and a phase $e^{i \theta}$ ($0 \le \theta \le 2 \pi$). Due to the phase of $e^{i \theta}$ there are an infinite number of equivalent vacua labeled by $\theta$. Usually one takes the arbitrary choice of 
$\theta =0$ as the vacuum for $\Phi$. This non-zero vacuum expectation value of the scalar field is responsible for giving masses to the $W^\pm$ and $Z^0$ gauge bosons of the $SU(2) \times U(1)$ Standard Model, while leaving the photon massless.  

Aside from the Standard Model, the Higgs mechanism has found application in the theory of superconductors via the Ginzburg-Landau model 
\cite{GL}. In the Ginzburg-Landau model the source of the non-zero order parameter/scalar field vacuum expectation value is due to the interaction between the electrons and the phonons of the background lattice. 

Another set of phenomena where a non-trivial vacuum is important are the Casimir effect \cite{casimir} and dynamical Casimir effect \cite{d-casimir}. In the canonical Casimir effect there are two, neutral, conducting plates which are placed a fixed distance apart. This restricts the type of quantum fluctuations that can occur between the plates as compared to outside the plates leading to an attractive force between the plates. In the dynamical Casimir effect the plates are moved with respect to one another and this results in the creation of photons out of the vacuum -- a result which has been observed relatively recently \cite{d-casimir1}.      

Below we will show that a massless scalar field placed in a gravitational wave background leads to the scalar
field developing a non-zero vacuum expectation value. We make a comparison of this gravitationally induced effect with 
the scalar field vacuum expectation value of spontaneous symmetry breaking as found in the Higgs mechanism and the Ginzburg-Landau model.
The comparison to the Ginzburg-Landau model is especially relevant since there the symmetry breaking is driven by the interactions induced by the phonons from the background lattice, whereas in the usual Higgs mechanism the symmetry breaking comes from the scalar field's self interaction. As in the Ginzburg-Landau model, here the scalar field's vacuum value is driven by interactions with the gravitational wave background. We also make a comparison between the present gravitationally induced vacuum expectation value and the dynamical Casimir. In the dynamical Casimir effect and the present case there is the possibility of producing {\it massless} particles from the vacuum. There are earlier works \cite{gibbons} \cite{garriga} which show that a plane gravitational wave background will not produce particles from the vacuum. We show how this is avoided exactly for the case of {\it massless} (scalar) fields.   

Finally, we connect the results of the present paper with other recent works that propose there is a shift of the pre-existing Higgs vacuum expectation value of the Standard Model either via stationary gravitational fields \cite{onifrio1, onifrio2} or via a gravitational wave background \cite{kurkov}. There is also very recent work \cite{caldwell} which discusses the consequences of the interaction of a gravitational wave background with a time-dependent vacuum expectation value from a (non-Abelian) gauge field. 

\section{Scalar field in gravitational wave background}

\subsection{Approximate gravitational wave background}

The equation for a complex scalar field, $\varphi$, in a general gravitational background is 
\begin{equation}
\label{KGvacuum1}
\frac{1}{{\sqrt { -  det \left[ {g_{\mu \nu } } \right]} }}\left( {\partial _\mu  g^{\mu \nu } \sqrt { - det \left[ {g_{\mu \nu } } \right]} \partial _\nu  } \right)\varphi  = 0.
\end{equation}

\noindent Following \cite{Jones16} we take the gravitational wave to travel in the positive $z$ direction and have the $+$ polarization. For this situation the metric \cite{Schutz} can be written as,
\begin{equation}
\label{metric}
ds^2 = -dt^2 + dz^2 + f(t-z)^2 dx^2 + g(t-z)^2 dy^2 =  du dv + f(u)^2 dx^2 + g(u)^2 dy^2,
\end{equation}
where in the last step we have switched to light front coordinates $u = z - t$ and $v = z+t$ with metric components $g_{uv} = g_{vu} = \frac{1}{2}$ and $g_{xx} = f(u) ^2$ and $g_{yy} = g(u)^2$. The metric functions $f(u)$ and $g(u)$ will be taken to be oscillatory functions of $u$. The determinant term in  \eqref{KGvacuum1} is  
$\sqrt{-det {\left[ {g_{\mu \nu } } \right]}} = \frac{fg}{2}$. Substituting the light front version of the metric into  equation \eqref{KGvacuum1} gives

\begin{equation}
\left( {4 f^2 g^2 \partial _u \partial _v  + 2 fg {\partial _u (fg)} \partial _v  +  g^2 \partial _x^2  +  f^2 \partial _y^2 } \right)\varphi  = 0.
\label{KGvacuum4}
\end{equation} 

\noindent We take the metric ansatz functions of the form $f\left( u \right) = 1 + \varepsilon \left( {u} \right)$, and $g\left( u \right) = 1 - \varepsilon \left( {u} \right)$ and substitute these into equation \eqref{KGvacuum4}. This form for $f\left( u \right)$ and $g\left( u \right)$ describes a wave propagating in the $z$ direction so that $x$ and $y$ directions should
be physically identical. Thus we require of the solution that $\left( {\partial _y^2 - \partial _x^2 } \right)\varphi=0$. Using this equation \eqref{KGvacuum4} becomes,

\begin{widetext}
\begin{equation}
\left[ {4 \left( {1 - 2\varepsilon ^2  + \varepsilon ^4 } \right)\partial _u \partial _v  - 4 \left( {1 - \varepsilon ^2 } \right)\varepsilon \left( {\partial _u \varepsilon } \right)\,\partial _v  + 
(1 + \varepsilon ^2 ) (\partial _x^2  + \partial _y^2 )} \right]\varphi  = 0.
\label{KGvacuum7}
\end{equation}
\end{widetext}

\noindent Finally we consider a sinusoidal, plane gravitational wave by taking $\varepsilon \left( u \right) = h_ +  e^{i Ku}$, where $h_{+}$ is a dimensionless amplitude and $K$ is a wave number. The metric must be real so it is understood that the metric components are obtained by taking the real part of the ansatz functions so that $f(u), g(u) =1 \pm h_+ e^{i Ku} \rightarrow 1 \pm h_+ \cos(K u)$. This real form still satisfies the linearized general relativistic field equations to which $f(u), g(u)$ are solutions. Substituting this choice of $\varepsilon (u)$ into equation \eqref{KGvacuum7} gives

\begin{equation}
 \left( {4 F(u) \partial _u \partial _v  - 4iKG(u)\,\partial _v  + H(u) (\partial _x^2  + \partial _y^2) }\right)\varphi  = 0,
\label{KGvacuum10}
\end{equation}

\noindent where $F\left( {u} \right)  \equiv  \left( {1 - 2h_ + ^2 e^{2iKu}  + h_ + ^4 e^{4iKu} } \right)$,  
$G\left( {u} \right)  \equiv  \left( {h_ + ^2 e^{2iKu}  - h_ + ^4 e^{4iKu} } \right)$, and $H(u) = \left( 1 + h_+^2 e^{2iKu} \right)$ . We separate equation \eqref{KGvacuum10} using 
$\varphi  = X\left( x \right)Y\left( y \right) U \left( u \right) V \left( v \right)$. The eigenvalue equations and associated solutions 
for $X(x)$ and $Y(y)$ are

\begin{equation}
 \partial _x^2 X =  - p^2 X \to X(x) = e^{ip x} ~~~~,~~~~  
 \partial _y^2 Y =  - p^2 Y \to Y(y) = e^{ip y}.  
\label{XYequations}
\end{equation}

The function $X(x)$ and $Y(y)$ are simply free waves as is to be expected since the gravitational wave is traveling 
in the $u=z-t$ direction, and $p$ is the common momentum in the $x, y$ directions. The common momentum in the $x$ and $y$ directions comes from the assumed symmetry in these transverse directions, and it also realizes the condition $\left( {\partial _y^2 - \partial _x^2 } \right)\varphi=0$ which we took above. Using \eqref{XYequations} we find that \eqref{KGvacuum10} becomes

\begin{equation}
F(u) \frac{{\partial _u U}}{U}\frac{{\partial _v V}}{V} - iKG(u)\frac{{\partial _v V}}{V} -  H(u) \frac{p^2}{2}  = 0.
\label{ASequation}
\end{equation}

\noindent Since the light front coordinate $v$ is orthogonal to $u$ and since the gravitational wave only depends
on $u$ one expects that the eigenfunction $V(v)$ also is solved by a free, plane wave, as was the case for
$X(x)$ and $Y(y)$. This is the case and we find

\begin{equation}
 - i\partial _v V = p_v V \to V(v) = e^{ip_v v} .
 \label{eigenvalueV}
\end{equation}

\noindent Substituting equation \eqref{eigenvalueV} into equation \eqref{ASequation} and defining $\lambda \equiv \frac{p ^2}{2 p_v}$ yields

\begin{equation}
i\frac{{\partial _u U(u)}}{U(u)} = \lambda \frac{H(u)}{F(u)} - K\frac{G(u)}{F(u)} ~.
 \label{eigenvalueU3}
\end{equation}

\noindent The term $i\frac{{\partial _u U(u)}}{U(u)}$ in \eqref{eigenvalueU3} represents the kinetic energy of the scalar field; the term $\lambda \frac{H(u)}{F(u)}$ represents
the interaction of the scalar field, via $\lambda$, with the gravitational background, via $\frac{H(u)}{F(u)}$; the term $K\frac{G(u)}{F(u)}$ represents a pure gravitational
potential term. Equation \eqref{eigenvalueU3} can be integrated to give,

\begin{equation}
U(u) = A e^{\frac{\lambda }{K}} e^{  \frac{- \lambda }{{K\left( {1 - h_ + ^2 e^{2iKu} } \right)}}} 
  \left( {1 - h_ + ^2 e^{2iKu} } \right)^{\frac{1}{2}\left( {\frac{\lambda }{K} - 1} \right)} e^{ - i\lambda u} ~,
 \label{eigenvalueU4}
\end{equation}

\noindent where $A e^{\frac{\lambda }{K}}$ is constant. The factor $e^{\frac{\lambda }{K}}$ was chosen to ensure that the eigenfunction for the $u$ direction becomes a free plane wave, $e^{- i\lambda u}$, as $h_+ \to 0$ ({\it i.e.} as the gravitational wave is turned off). Collecting together all the solutions in $x, y, v$ and $u$ directions gives the solution of the scalar field in the gravitational background,

\begin{equation}
\varphi  = A e^{\frac{\lambda }{K}} e^{ - \frac{\lambda }{{K\left( {1 - h_ + ^2 e^{2iKu} } \right)}}} \left( {1 - h_ + ^2 e^{2iKu} } \right)^{\frac{1}{2}\left( {\frac{\lambda }{K} - 1} \right)} 
e^{ - i\lambda u} e^{ip_v v} e^{ip x} e^{ip y} + B .
\label{Sfield}
\end{equation}

\noindent $A$ is a normalization constant and we have added a constant $B$ which is allowed by the shift symmetry of solutions to  \eqref{KGvacuum4}. Below we choose $B= -A$. This choice of $B$ ensures that if one turns off the gravitational background ($h_+ \to 0$) and also takes the field momenta to zero ($\lambda, p_v, p \to 0$) then $\varphi \to 0$. This solution for the scalar field given in \eqref{Sfield} is very similar to the form of the solution found in \cite{Padmanabhan99} for the {\it static} electric field evaluated in light front coordinates. Here we have a massless scalar field in a gravitational wave background.  

In the limit  $h_+ \to 0$ \eqref{Sfield} goes to the expected flat space-time solution of a free wave scalar field,
$\varphi _0  \propto e^{ - i\lambda u} e^{ip_v v} e^{ip x} e^{ip y} \to \varphi_0 \propto  e^{i(p_v + \lambda)t} e^{i(p_v -\lambda) z} 
e^{ip x} e^{ip y}$. The second version of $\varphi _0$ is the conversion from light front coordinates back to $t ,z$ coordinates. Defining 
an energy $E_0 = p_v + \lambda$ and a momentum in the z-direction $p_z = p_v -\lambda$ one sees that the usual dispersion relationship
for a massless field ({\it i.e.} $E_0 ^2 = p_x^2 + p_y^2 + p_z^2 \to 2p^2 + p_z^2$ is recovered if one recalls that $\lambda = \frac{p^2}{2 p_v}$). Next, taking the limit of \eqref{Sfield} when all the transverse momenta of the scalar field go to zero ({\it i.e.} $p_x=p_y=p \to 0$ and $\lambda \rightarrow 0$) one finds

\begin{eqnarray}
\label{phi-higgs}
\varphi (u, v)  &=& \mathop {\lim }\limits_{p_v \to 0} A \left( {1 - h_ + ^2 e^{2iKu} } \right)^{-\frac{1}{2}} e^{i p_v v} - A \nonumber \\
& & \to  A \left[ \left( {1 - h_ + ^2 e^{2iKu} } \right)^{-\frac{1}{2}} - 1 \right] \approx \frac{A}{2} h_+ ^2 e^{2i Ku}+ \frac{3A}{8}  h_+ ^4 e^{4i Ku}~.
\end{eqnarray}

\noindent In \eqref{phi-higgs} we have taken the limit $p_v \to 0$ after the limit of the transverse momenta going to zero
$p_x=p_y=p \rightarrow 0$. Using the vacuum solution in \eqref{phi-higgs} the magnitude square of $\varphi$ is

\begin{equation}
\varphi ^* \varphi \approx \frac{A^2 h_+ ^4}{4} + \frac{3 A^2 h_+^6}{8} \cos(2 K u ).
\label{phi-higgs-2}
\end{equation}

\noindent Equations \eqref{phi-higgs} and \eqref{phi-higgs-2} show that in the presence of a gravitational wave background, that in addition to the vacuum state solution $\varphi = 0$ ($\varphi=0$ is always a solution to \eqref{KGvacuum1}), there are also the vacuum solutions given by $\varphi (u,v)$ in \eqref{phi-higgs}. A common feature shared by the present example and the usual scalar field symmetry breaking is that the scalar field magnitude only depends on the parameters of the interaction -- $\varphi^* \varphi$ depends on $h_+$ and $K$ (the amplitude  and wave number of the gravitational wave background) while for the canonical Higgs field example of \eqref{higgs} $\Phi^* \Phi$ depends on $m^2$ and $\Lambda$ (the parameters of the scalar fields self interaction).  

One can compare the above results with the Casimir effect \cite{casimir} and the dynamical Casimir effect \cite{d-casimir}.
Both of the Casimir and dynamical Casimir effects involve a non-trivial vacuum state due to the restriction of the quantum fluctuations of fields placed between two neutral conducting, infinite plates. And in the dynamical Casimir effect the time dependent oscillation of the plates creates photons out of the vacuum. In the present case our interpretation of the scalar field solution in \eqref{phi-higgs} is that the time dependent oscillations of the gravitational field create scalar field quanta out of the vacuum.   

In support of this interpretation, that the scalar field solution from \eqref{phi-higgs} represents the creation of scalar field quanta by the gravitational wave background, we look at the current in the $u$-direction associated with $\varphi$ from \eqref{phi-higgs} which to lowest order in $h_+$ is given by

\begin{equation}
\label{current-u}
j_u = -i (\varphi ^* \partial _u \varphi - \varphi \partial _u \varphi ^*) \approx A^2 h_+^4 K + \frac{9}{4} A^2 h_+ ^6 K \cos(2 K u) ~,
\end{equation}

\noindent If we take the lowest order of the current in \eqref{current-u} (or time averaging $j_u$ in \eqref{current-u}) we find a constant scalar field current in the $u$ direction of magnitude $A^2 h_+^4 K$. Our interpretation of this result is that the incoming gravitational wave current in the $u$ direction creates an outgoing scalar field current given, to leading order, by the first term in \eqref{current-u}. This picture is further supported by looking at the tree-level, Feynman diagram processes of $graviton + graviton \to photon + photon$, where in our case $photon$ is really the massless scalar field quanta. In reference \cite{skobelev} this tree level process was calculated and found to be non-zero in general. In particular it is non-zero when the incoming gravitons and outgoing photons travel in the same direction . In this work we are looking at collections of gravitons and scalar field quanta ({\it i.e.} gravitational plane waves and scalar field plane waves). Thus, if the process $graviton + graviton \to photon + photon$ is non-zero at the individual quanta level, as shown in \cite{skobelev}, then this implies it should be non-zero for a collection of these quanta {\it i.e.} gravitational and scalar field plane waves.  

To conclude this subsection we recall that there are well known restrictions against the creation of field quanta by a plane gravitational wave background \cite{gibbons} \cite{garriga}, which is what we are proposing above. However in reference \cite{gibbons} a loop hole was given -- particle creation might occur if the scalar field were massless and if the momenta of the scalar field quanta were in the same direction as the gravitational plane wave. These are exactly the conditions we have here -- the scalar field is massless and is traveling in the same direction as the gravitational wave since from \eqref{phi-higgs} $\varphi$ only depends on $u=z-t$. In more detail it was shown in \cite{garriga} that the Bogoliubov $\beta$ coefficients, which indicate particle production, were proportonal to energy-momentum conserving delta functions
\begin{equation}
\label{b-beta}
\beta_{ij} = \langle u_i ^{out} | u_j ^{in~*} \rangle \propto \delta (k_{-} + l_{-}) ~,
\end{equation}
where $k_{-} = \frac{\omega - k_z}{2}$ and $l_{-} = \frac{\omega - l_z}{2}$ are the light front momenta of the scalar field before and after \footnote{In \cite{garriga} as well as in \cite{gibbons} a sandwich gravitational wave background was considered. The plane gravitational wave background of \eqref{metric} was sandwiched, before and after, by Minkowski space-time. The functions $u_i ^{in}$ and $u_j ^{out}$ are the solutions in the asymptotic Minkowski regions that are connected to each other through the intermediate plane wave gravitational background.}, 
$\omega =\sqrt{{\bf k}^2 + m^2}$ or $\omega =\sqrt{{\bf l}^2 + m^2}$ respectively, and the indices $i,j$ label the momenta of the outgoing and ingoing scalar field quanta. If $m \ne 0$ it is easy to see that $k_{-} + l_{-}$ can not vanish. If however, as is true in the case consider here, $m=0$ and ${\bf k, l} \to k_z, l_z$ ({\it i.e.} the before and after momemta of the scalar field is purely along the $+z$ direction) then $k_{-} + l_{-}$ vanishes and the Bogoliubov $\beta$ coefficient is non-zero indicating particle production. 

\subsection{Exact gravitational wave background}
    
One might ask to what extent the linear approximation for the gravitational wave -- namely that $f(u) = 1 + \epsilon (u)$ and
$g(u) = 1- \epsilon (u)$ with $\epsilon (u) = h_+ e^{iKu}$ -- is crucial in obtaining the result in \eqref{phi-higgs}. What if one
took an exact, gravitational plane wave solution instead of a linearized approximation?  To this end we now repeat briefly the above analysis for an exact solution for the plane wave metric in the $+$ polarization. The ansatz functions $f(u)$ and $g(u)$ will be exact, plane wave solutions if they satisfy the general relativistic field equations in this case which are of the form ${\ddot f}/ f + {\ddot g}/ g = 0$ \cite{Schutz}. One simple exact, plane wave, solution is $f(u) = e^{-iKu} e^{ - Ku}$ and $g(u) = e^{-iKu} e^{Ku} $. These ansatz functions have plane wave parts ($e^{-iKu}$) but they also have exponentially growing or decaying amplitudes ($e^{ \pm Ku}$). Near $u=0$ one has oscillating, wave solutions due to the $e^{-iKu}$ parts of the ansatz function, but as $u$ moves away from $u=0$ the $e^{\pm Ku}$ terms act like growing/decaying  amplitudes. Because of this these solutions can only be of use for a restricted range of $u$ near $u=0$. Asymptotically, as $u\rightarrow \infty$, the functions $f(u), g(u)$ are not physically acceptable. In this case we are dealing with an exact solution to the non-linear general relativistic field equations so one may ask if taking the real part of the complex form of the ansatz functions will still be a solution to ${\ddot f}/ f + {\ddot g}/ g = 0$ due to the non-linearity of general relativity. One can show that taking the real part of the ansatz functions ({\it i.e.} $f(u) = e^{-iKu} e^{ - Ku} \to \cos(Ku) e^{- Ku}$ and $g(u) = e^{-iKu} e^{Ku} \to \cos(Ku) e^{Ku}$) is still a solution to ${\ddot f}/ f + {\ddot g}/ g = 0$. However, as in the previous linearized case, it is much easier to work with the complex form of the ansatz functions when one uses the background metric in the equation for the complex scalar field. 

Using the above metric background we substitute $f(u) = e^{-iKu} e^{ - Ku}$ and $g(u) = e^{-iKu} e^{Ku}$ into equation \eqref{KGvacuum4},

\begin{widetext}
\begin{equation}
\left( {4 e^{-4iKu} \partial _u \partial _v  + 2 e^{-2iKu} \partial _u \left( {e^{-2iKu} } \right)\partial _v  + e^{-2iKu} e^{2Ku} \partial _x^2  + e^{-2iKu} e^{ - 2Ku} \partial _y^2 } \right)\varphi  = 0,
\label{Exact1}
\end{equation}
\end{widetext}

\noindent and making the substitution $\varphi  = U(u) V(v) X(x) Y(y) = U(u) e^{ip_v v} e^{ip x} e^{ip y}$, we find

\begin{equation}
\left( i \frac{{\partial _u U}}{U} +  K - \lambda e^{  2iKu} \cosh(2 K u) \right) = 0 ~,
\label{Exact2}
\end{equation}

\noindent where $\lambda =\frac{p^2}{2p_v}$ as before. Equation \eqref{Exact2} can be compared to \eqref{eigenvalueU3} in the sense that $i \frac{{\partial _u U}}{U}$ is the kinetic energy term of the scalar field, $\lambda e^{  2iKu} \cosh(2 K u)$ is an interaction between the scalar field and the gravitational background, and $K$ is a pure gravitational potential term. 

In the limit when the gravitational wave is absent ({\it i.e.} $K \to 0$) the solution to \eqref{Exact2} is again given by 
$\varphi _0  \propto e^{ - i\lambda u} e^{ip_v v} e^{ip x} e^{ip y}$. Restoring the gravitational background ({\it i.e.} $K \ne 0$) the solution to \eqref{Exact2} is $U (u) = A e^{\left( {\frac{{\left( {-1 - i}  \right)\lambda}}{{8K }} e^{  2iKu} e^{2Ku}  + \frac{{\left( {-1 + i}  \right)\lambda}}{{8K}} e^{  2iKu} e^{ - 2Ku} } \right)} e^{iKu}$, where $A$ is a constant similar to that found in \eqref{eigenvalueU4} and the scalar field takes the form

\begin{equation}
\varphi (u, v, x, y) = A e^{\left( {\frac{{\left( {-1 - i}  \right)\lambda}}{{8K }} e^{  2iKu} e^{2Ku}  + \frac{{\left( {-1 + i}  \right)\lambda}}{{8K}} e^{  2iKu} e^{ - 2Ku} } \right)} e^{iKu}
e^{ip_v v} e^{ip x} e^{ip y} + B,
\label{Exact4}
\end{equation}

\noindent where $A$ is a normalization constant. We have again added a constant $B$ via the shift symmetry of solutions to \eqref{KGvacuum4}. As before we set $B=-A$ so that $\varphi (u,v,x,y) \to 0$ when the gravitational wave background is turned off and when the scalar field momenta go to zero. Here we do not have an $h_+$ since the ``amplitude" is given  by the $e^{\pm Ku}$ terms in $f(u), g(u)$. As before if we take the limit of the massless scalar field to its vacuum state by taking its energy and momenta parameters to zero  ({\it i.e.} taking the limit $p_x=p_y=p   \to 0$ and  $\lambda \to 0$) one finds that as before $\varphi$ and $\varphi \varphi^*$ do not go to zero but rather 
\begin{equation}
\label{phi-exact}
\varphi = \mathop {\lim }\limits_{p_v \to 0} A e^{iKu} e^{i p_v v} = A  (e^{iKu} -1) ~~~~;~~~~ \varphi ^* \varphi = 2 A^2  (1- \cos(K u))~.
\end{equation} 
In \eqref{phi-exact} we have again taken the limit that $p_v$ becomes arbitrarily small ($p_v \rightarrow 0$). With this we see that 
$\varphi$ depends on the gravitational wave background through the wave number, $K$. For this exact solution metric we again find that the scalar field acquires a non-zero, space-time dependent vacuum value even when one takes the limit of all the scalar field momenta going to zero. Also as in the previous subsection we find that $\varphi ^* \varphi$ has a constant term (the $2 A^2$ term in \eqref{phi-exact} which corresponds to the $\frac{A^2 h_+ ^4}{4}$ term in \eqref{phi-higgs-2}) and a space-time dependent part (the $-2 A^2 \cos(Ku)$ term in \eqref{phi-exact} which corresponds to the $\frac{3 A^2 h_+^6}{8} \cos(2 K u )$ term in \eqref{phi-higgs-2}). As for the plane wave solution of the previous subsection we can calculate the current in the $u$-direction and find
\begin{equation}
\label{current-u-2}
j_u = -i (\varphi ^* \partial _u \varphi - \varphi \partial _u \varphi ^*) = 2 A^2 K (1 - \cos(Ku)) ~.
\end{equation}  
The current above is similar to the one found in the previous subsection in \eqref{current-u} -- there is a constant term, $2 A^2 K$, and a space-time dependent term $-2 A^2 K\cos(Ku)$. These can be compared to the terms $A^2 h_+^4 K$ and $\frac{9}{4} A^2 h_+ ^6 K \cos(2 K u)$ in \eqref{current-u}. If one time averages the current in \eqref{current-u-2} one finds $\langle j_u \rangle = 2 A^2 K$ {\it i.e.} one has a constant current in the $u$ direction. We take the same interpretation of $\langle j_u \rangle$ as the leading term of $j_u$ from \eqref{current-u} -- $\langle j_u \rangle$ represents a scalar field plane wave traveling in the $u$-direction produced by the initial gravitational wave.  

\section{Discussion and Conclusions}

We have shown that a massless scalar field placed in a plane, gravitational wave background will develop a space-time dependent, non-zero vacuum value given by \eqref{phi-higgs} \eqref{phi-higgs-2} even in the limit when all the momentum parameters of the scalar field are taken to zero ({\it i.e.} $p_v, p , \lambda \rightarrow 0$). This is different from what happens to the massless scalar field solution in Minkowski space-time, where when one takes the zero energy-momentum limit the scalar field vanishes. We have drawn attention to the similarity of this gravitationally induced scalar field vacuum value with the vacuum expectation value in the Higgs phenomenon and with the dynamical Casimir effect. 

Three potential physical consequences of this gravitationally induced vacuum value for the scalar field are: (i) the production of massless field quanta, such as photons, from the gravitational wave background; (ii) the usual Higgs vacuum expectation value of the Standard Model may be modified or even generated by stationary and/or time dependent gravitational backgrounds; (iii) the interplay between gravitational waves and scalar and gauge fields in the early Universe may lead to observational consequences at the present time. 

Point (i) was investigated in references \cite{Jones16, Jones15}. Here we calculated the currents connected with the scalar field (see equations \eqref{current-u} and \eqref{current-u-2}) that the time averaged current, $\langle j_u \rangle$ was a constant which we interpreted  as the incoming gravitational wave creating an outgoing scalar field. There is a well known restriction against the creation of fields from an incident, plane gravitational plane wave \cite{gibbons}. The present work avoids this conclusion by using the loop hole mentioned in \cite{gibbons} that the prohibition only applied to massive fields. Here we considered a massless field that travels in the same direction as the initial gravitational wave. Furthermore in \eqref{b-beta} we have shown that the the Bogoliubov coefficients indicating  
particle production, as calculated in \cite{garriga} for a gravitational plane wave sandwich background, are non-zero exactly in the limit $m \to 0$ and the momenta of the field quanta being in the same direction as the gravitational wave ({\it i.e.} exactly the conditions of this work). The production of fields by an incoming gravitational plane wave background is also in agreement with the non-zero, tree-level amplitudes for $graviton + graviton \to photon + photon$ coming from Feynman diagram calculations \cite{Calmet16, skobelev, bohr}. Also since these processes occur at the tree-level Feynman diagrams they are in some sense classical effects.   

Point (ii) was discussed in references \cite{onifrio1, onifrio2, kurkov}. In the works \cite{onifrio1, onifrio2} the idea was considered that the usual Higgs expectation value can be shifted by the effect of a static or stationary gravitational background ({\it e.g.} the Schwarzschild or Kerr space-times). This effect requires a coupling between the scalar field and the gravitational field of the form $\xi \phi R$ or $\xi \phi K$ where $\xi$ is the coupling and $R$ is the Ricci scalar and $K=R_{\alpha \beta \gamma \delta}
R^{\alpha \beta \gamma \delta}$ is the Kretcschmann scalar. This coupling of the scalar field to the gravitational field results 
in a shifting of the pre-existing Higgs vacuum expectation value with an associated shift in particle masses. This shift is potentially observable. In the work \cite{kurkov} the idea of a scalar field-gravitational background coupling of the form $\xi \phi R$ is again considered but now the gravitational background is both space and time dependent. Again it is found that the gravitational background can shift the Higgs vacuum expectation value, but now this shift is space-time dependent as is the case for the results of the present work. Further it was found in \cite{kurkov} that even when there is no vacuum expectation value of the scalar field, one can be generated from the interaction with the gravitational background as is the case for our results.   

Point (iii) was very recently proposed in reference \cite{caldwell} where the interplay of a gravitational wave background with a cosmological 
non-Abelian gauge field was considered. The non-Abelian gauge field was assumed to have a {\it pre-existing} vacuum expectation value of the form $A_i ^a = \phi (\tau) \delta _i ^a$ where $\phi (\tau)$ is a scalar function of the proper time $\tau$, and the indices $i$ and $a$ were space and ``color" indices respectively. This interplay between the gravitational wave and the pre-existing, non-Abelian gauge field might lead to interesting and potentially observable phenomenon such as neutrino-like oscillations between the gravitational field and the non-Abelian gauge field. In contrast, in the present work, a space-time dependent scalar field, as given in \eqref{phi-higgs}, is generated out of the vacuum by the gravitational wave background. Similarly one might conjecture that the phenomenon proposed in \cite{caldwell} could work with the gravitational wave background generating a non-zero vacuum expectation value for the gauge field from the vacuum rather than requiring a pre-existing gauge field.       

Finally we want to point out that, like the standard Higgs mechanism of particle physics, the present generation of the vacuum expectation value of the scalar field by the gravitational wave background is already implied at the classical level. In the usual Higgs mechanism, as given in \eqref{higgs}, the non-zero vacuum expectation value of $\Phi_0 = \sqrt{\frac{-m^2}{\Lambda}}$ is obtained from the classical, scalar field Lagrangian \eqref{higgs}. In a similar way the non-zero vacuum value for $\varphi$ from the gravitational wave background already emerges by examining the system of a classical scalar field interacting with a classical gravitational background as given in \eqref{phi-higgs} and \eqref{phi-exact}. Also the view that the leading term in the scalar field currents, $j_u$, from \eqref{current-u} and \eqref{current-u-2} represented production of the scalar field from the incoming gravitational wave, finds support from the non-zero, tree-level Feynman diagram process $graviton + graviton \to photon + photon$. This further indicates the classical nature of the scalar field vacuum expectation values found here, since tree-level Feynman diagrams represent the classical limit of a given interaction.     

{\bf Acknowledgment}

DS is supported by grant $\Phi.0755$ in fundamental research in Natural Sciences by the Ministry of Education and Science of Kazakhstan.

\end{document}